\documentclass[a4paper,11pt]{article}
\usepackage[german,english]{babel}
\usepackage[T1]{fontenc}
\usepackage{amssymb}
\usepackage{amsmath}
\usepackage{amsfonts}
\usepackage[mathscr]{euscript}
\usepackage{aeguill} 
\newcommand{\artanh}{\mathop\mathrm{artanh}\nolimits}

\begin{document}

\title{Pekeris-type approximation for the $l$-wave in a Pöschl-Teller potential} 
\author{Stoian I. Zlatev\thanks{Email address: zlatev@ufs.br}\\
Departamento de Física - CCET, \\
Universidade Federal de Sergipe\\
Av. Marechal Rondon, s/n, Jardim Rosa Elze \\
CEP 49100-000 - São Cristóvão/SE, Brasil}

\maketitle

\begin{abstract}
An approximation for the centrifugal term which transforms the radial equation for a particle in a Pöschl-Teller potential into an exactly solvable approximate equation is proposed. In contrast to the approximations known in the literature, the new one can be used with the most general form of the central Pöschl-Teller potential.  
An approximate expression for the bound-state spectrum is obtained. 
\end{abstract}



\maketitle

\section{Introduction}
The Pöschl-Teller functions \cite{poschl-teller33} are among the one-dimensional potentials for which the discrete spectrum of the quantum-mechanical Hamiltonian is known, as well as the eigenfunctions  \cite{infeld-hull51}.  Attracting less attention than Morse potential or Hulburt-Hirschfelder potential,
the second Pöschl-Teller function has been successfully used for the description of the vibrational and rotational motion of diatomic molecules \cite{davies49,beckel57}. 

A standard separation of variables reduces  the bound-state problem for a particle in a Pöschl-Teller potential to an one-dimensional eigenvalue problem but the exact solutions of the latter are not known for non-zero values of the azimuthal quantum number $l$. A suitable approximation for the centrifugal term in the radial equation transforms the eigenvalue problem into one that can be solved analytically \cite{wei-dong10,xu+10,ikhdair-hamzavi12}. The approximations used are the Greene-Aldrich one \cite{greene-aldrich76} or its modifications \cite{jia+09a,you+13}. They lead to exactly solvable approximate equations only if one of the parameters in the Pöschl-Teller function, namely the parameter $r_{0}$ in eq.~(\ref{2pt-potential}),  is set equal to zero. 

The aim of the present letter is to propose a new  approximation for the centrifugal term that leads to an exactly solvable one-dimensional eigenvalue problem  for the most general form of the Pöschl-Teller potential. 

\section{The Scrödinger equation with a Pöschl-Teller potential}
In the Born-Oppenheimer
 approximation, the vibrational and rotational motion of a diatomic molecule is governed by a one-particle Hamiltonian 
\[
 H= -\frac{\hbar^{2}}{2M}\nabla^{2}+ V(r),
\]
where $M$ is the system reduced mass and the central  potential $V(r)$ satisfies some general requirements listed by Morse \cite{morse29}.  The second Pöschl-Teller potential \cite{poschl-teller33},
\begin{equation}
\label{2pt-potential}
V(r)= 
\frac{A}{\sinh^{2} \alpha(r-r_{0})}- \frac{B}{\cosh^{2}\alpha(r-r_{0})},
\end{equation}
where the parameters $A$, $B$, $\alpha$ are positive and $r_{0}$ is non-negative, satisfies all the  requirements if $A<B$, which we assume in the sequel.  
In this case, the potential has a minimum at 
\begin{equation}
 \label{re}
 r_{e}= r_{0}+\frac{1}{\alpha}\artanh\sqrt[4]{\frac{A}{B}}.
\end{equation}
The radial equation, obtained after the separation of the variables 
$\psi(r,\theta, \phi)= r^{-1}R(r)Y(\theta,\phi)$,
can be written as 
\begin{equation}
 \label{radial}
 \frac{d^{2}R}{dr^{2}}+ \left[-\alpha^{2}\left(\frac{\nu(\nu-1)}{\sinh^{2} \alpha(r-r_{0})}- \frac{\mu(\mu+1)}{\cosh^{2}\alpha(r-r_{0})}\right)-\frac{l(l+1)}{r^{2}}+ \epsilon \right]R=0,
\end{equation}
where $\epsilon$ is a spectral parameter related to the energy in an obvious way,
$l$ is the azimuthal  quantum number and the parameters $A$ and $B$ have been changed for the positive solutions $\nu$ and $\mu$ of 
the equations
\begin{equation}
\label{numu}
\alpha^{2} \nu(\nu-1)= \frac{2M}{\hbar^{2}}A, \qquad \alpha^{2} \mu(\mu+1)=\frac{2M}{\hbar^{2}} B.
\end{equation}
For $l=0$, the square-integrable  solutions (on the interval $r>r_{0}$) of eq.~(\ref{radial})  are known  and can be expressed in terms of elementary functions \cite{infeld-hull51}. 
The number of the eigenvalues is  finite and they are given by 
\begin{equation}
 \label{epsilonn}
 \epsilon_{n}= -\alpha^{2}(\mu-\nu-2n)^{2}, \qquad n=0,1,\dots <\frac{\mu-\nu}{2}.
\end{equation}
The exact eigenvalues for $l>0$ are not known,, neither is known the exact form of the corresponding eigenfunctions. Therefore, it is natural to look for a suitable approximating function for the centrifugal term 
\begin{equation}
 \label{centrifugal}
 \frac{l(l+1)}{r^{2}}= \frac{l(l+1)}{r_{e}^{2}}\left(\frac{r_{e}}{r}\right)^{2},
\end{equation}
which, replacing the centrifugal term in (\ref{radial}), yields an exactly solvable approximate equation.   

\section{Approximations for the centrifugal term used in the case $r_{0}=0$}
Such approximations have been proposed \cite{xu+10,wei-dong10,ikhdair-hamzavi12,you+13} for the case when the adjustable parameter $r_{0}$ in the Pöshl-Teller function is set equal to zero. The approximation
\begin{equation}
 \label{greene-aldrich}
 \frac{1}{r^{2}}\approx  F_{1}(r, \alpha)=4\alpha^{2}\frac{e^{-2\alpha r}}{\left(1-e^{-2\alpha r}\right)^{2}}= \frac{\alpha^{2}}{\sinh^{2}(\alpha r)}
\end{equation}
introduced by Greene and Aldrich \cite{greene-aldrich76} has been used in Ref.~\cite{wei-dong10} while  an improved approximation \cite{jia+09a}
\begin{equation}
 \label{ga-improved}
 \frac{1}{r^{2}}\approx F_{2}(r, \alpha)=4\alpha^{2}\left[\frac{1}{12}+\frac{e^{-2\alpha r}}{\left(1-e^{-2\alpha r}\right)^{2}}\right]=
 \frac{\alpha^{2}}{\sinh^{2}(\alpha r)}+\frac{\alpha^{2}}{3}
\end{equation}
has been used in Refs.~\cite{xu+10,ikhdair-hamzavi12}. 
Both the approximations, and, especially, the improved one, are good ``for small values of the parameter $\alpha$'' \cite{jia+09}, or, in other words, they are good approximations for the function $1/r^{2}$ in the region $r\lessapprox\alpha^{-1}$. In the range where the centrifugal term ``has any appreciable effect'' \cite{morse29}, $r-r_{e}$ is small, but not necessarily $r$. 

In the improved approximation (\ref{ga-improved}), a quantity corresponding to the constant term in the Laurent expansion 
\[
 \frac{1}{\sinh^{2}{z}}= \frac{1}{z^{2}}-\frac{1}{3}+ \dots
\]
is subtracted. This results in an improvement for small values of $r$ at the price of a discrepancy for large values of $r$ since $F_{2}(r,\alpha)\to \alpha^{2}/3\not= 0$ when $r$ goes to infinity.  

The third approximation \cite{you+13} 
\begin{equation}
 \label{thirdappr}
 \frac{1}{r^{2}}\approx F_{3}(r, \alpha)=
 \frac{\alpha^{2}}{\sinh^{2}(\alpha r)}+\frac{t\alpha^{2}}{\cosh^{2}(\alpha r)}
\end{equation}
replaces the $r$-independent term in $F_{2}$ by $t\alpha^{2}/\cosh^{2}(\alpha r)$, where $t$ is a parameter. For $t=1/3$, the function $F_{3}$ shows the same behavior as $F_{2}$ for small values of $r$, while the correct limit at infinity is recovered. 

These approximations are useful if $r_{0}=0$ in the expression (\ref{2pt-potential}). However, the general form of the potential (\ref{2pt-potential}) is also of interest. For a large number of diatomic molecules, for example, the appropriate value of the parameter $r_{0}$ is different from zero \cite{beckel57}. 

\section{A Pekeris-type approximation for the centrifugal term}

The Pekeris approximation \cite{pekeris34} for the centrifugal term has been successfully applied to the Schrödinger equation with the Morse potential. It has been shown recently \cite{wei-dong10b,wei-dong10a,ferreira-prudente13} that similar  schemes can be used with the  
Manning-Rosen and Rosen-Morse potentials.  

A Pekeris-type approximation is based \cite{pekeris34,wei-dong10b,wei-dong10a,ferreira-prudente13} on the expansion of the centrifugal term (\ref{centrifugal}) in  powers of $y-y_{e}$, where $y=f(r)$ is a suitable function of $r$ and $y_{e}=f(r_{e})$. The function $f$ is chosen in such a way that the 
replacement of the dimensionless quantity $(r_{e}/r)^{2}$ in the centrifugal term by the first three terms in the expansion 
\begin{equation}
 \label{pekeris-exp}
 \left(\frac{r}{r_{e}}\right)^{2}= 1+c_{1}(y-y_{e})+ c_{2}(y-y_{e})^ {2}+ \mathscr{O}\left((y-y_{e})^{3}\right)
\end{equation}
yields an exactly solvable approximate radial equation. The higher order terms can be treated as a perturbation \cite{pekeris34}. 

To obtain an approximation transforming eq.~(\ref{radial}) into an exactly solvable approximate equation for any value of the azimuthal quantum number $l$, one can use the linear independence of the terms in the Pöschl-Teller potential (\ref{2pt-potential}). Let us put
\[
 y= \frac{1}{\sinh^{2} \alpha(r-r_{0})}+\frac{C}{\cosh^{2}\alpha(r-r_{0})},
\]
where $C$ is a constant. While $y-y_{e}$ contains only terms proportional to $\sinh^{-2}\alpha(r-r_{0})$,  $\cosh^{-2}\alpha(r-r_{0})$, and 1, this  is not true regarding $(y-y_{e})^{2}$, for any value of $C$. However, for a certain value of $C$, the coefficient $c_{2}$ in (\ref{pekeris-exp}) vanishes. Indeed, 
\[
 c_{2}= \frac{1}{2}\,\frac{d^{2}}{dy^{2}}\left(\frac{r_{e}}{r}\right)^2\bigg|_{y=y_{e}}
\]
and the condition $c_{2}=0$ is equivalent to 
\begin{equation}
 \label{eqn-r}
 \left[r\frac{d^{2}y }{dr^{2}}+3\frac{dy}{dr}\right]\Bigg|_{r=r_{e}}=0
\end{equation}
as a simple calculation shows. Calculating the derivatives, substituting them into eq.~(\ref{eqn-r}), solving the equation obtained with respect to $C$, and taking into account eq.~(\ref{re}), one gets
\[
C= - \left(\frac{b}{a}\right)^{4}\frac{\alpha r_{e}(3b^{2}-a^{2})-3ab}
{\alpha r_{e}(3a^{2}-b^{2})-3ab}, 
\]
where  
\[
a= \sqrt[4]{\nu(\nu-1)}, \qquad b=\sqrt[4]{\mu(\mu+1)}. 
\]
Then, calculating 
\begin{align}
 \label{c1}
c_{1}=& \frac{d}{dy}\left(\frac{r_{e}}{r}\right)^{2}\Bigg|_{y_{e}}= 
\frac{a^{3}\left[\alpha r_{e}(b^{2}-3a^{2})+3ab\right]}{4(\alpha 
r_{e})^{2}b(b^{2}-a^{2})^2},\\
\label{ye}
y_{e}=&3\frac{(b^{2}-a^{2})^{2}}{a^{4}}\cdot \frac{\alpha 
r_{e}(a^{2}+b^{2})-ab}{\alpha r_{e}(b^{2}-3a^2)+3ab},
\end{align}
and substituting  it in (\ref{pekeris-exp}), one obtains
\begin{align*}
\left(\frac{r_{e}}{r}\right)^{2}=& 1+ \frac{3}{4}\cdot\frac{ab-\alpha 
r_{e}(a^{2}+b^{2})}{(\alpha r_{e})^{2}ab}\\
&+\frac{1}{4(\alpha r_{e})^{2}(b^{2}-a^{2})^{2}}\Big[
\frac{a^{3}}{b}\cdot
\frac{3ab+\alpha r_{e}(b^{2}-3a^{2})}{\sinh^{2}\alpha 
(r-r_{0})}\\
&-\frac{b^{3}}{a}\cdot\frac{3ab+\alpha r_{e}(a^{2}-3b^{2})}{\cosh^{2}\alpha 
(r-r_{0})}
\Big]+ \mathscr{O}\left((y-y_{e})^{3}\right).
\end{align*}

\section{The bound states}
Replacing the function $(r/r_{e})^{2}$ in the centrifugal term  by the sum of the first two terms
in the expansion (\ref{pekeris-exp}), one obtains the (approximate) equation
\begin{align}
 \notag
\frac{d^2R}{dr^{2}}&- \alpha^{2}
\left[
\frac{\nu(\nu-1)+\frac{c_{1}}{(\alpha r_{e})^{2}}l(l+1)}{\sinh^{2} \alpha(r-r_{0})}
- \frac{\mu(\mu+1)-\frac{c_{1}C}{(\alpha r_{e})^{2}}l(l+1)}{\cosh^{2}\alpha(r-r_{0})}
\right]R\\
\label{radial-2}
&+\left[\epsilon-\frac{l(l+1)}{r_{e}^{2}}(1-c_{1}y_{e})
\right] R=0,
\end{align}
which has the form of eq.~ (\ref{radial}) with $l=0$ and constants $\nu^{\prime}$, $\mu^{\prime}$, and $\epsilon^{\prime}$ given by 
\begin{align}
 \label{nuprime}
 &\nu^{\prime}(\nu^{\prime}-1)=\nu(\nu-1)+ \frac{c_{1}}{(\alpha r_{e})^{2}}, \\
\label{muprime}
 & \mu^{\prime}(\mu^{\prime}+1)=\mu(\mu+1)- \frac{c_{1}C}{(\alpha r_{e})^{2}} , \\
\label{epsilonprime}
 &\epsilon^{\prime}= \epsilon-\frac{l(l+1)}{r_{e}^{2}}(1-c_{1}y_{e}).
\end{align}
Solving these equations and substituting the values in the eq.~ (\ref{epsilonn}), one obtains that the eigenvalues $\epsilon_{nl}$ for the equation (\ref{radial}) with azimuthal  quantum number $l$ are (approximately) given by 
\begin{align}
\notag
 \epsilon_{nl}
 \label{epsilonnl}
 \notag
= &-\alpha^{2}\Bigg\{\Bigg[2\left(n+\frac{1}{2}\right)\\
\notag
&+\sqrt{\frac{1}{4}+a^{4}+\frac{a^{3}\left[3ab+\alpha r_{e}(b^{2}-3a^{2})\right]}{4(\alpha r_{e})^{4}b(b^{2}-a^{2})^{2}}l(l+1)}\\
\notag
&-\sqrt{\frac{1}{4}+b^{4}+\frac{b^{3}\left[3ab+\alpha r_{e}(a^2-3b^{2})\right]}{4(\alpha r_{e})^{4}a(b^{2}-a^{2})^{2}}l(l+1)}
 \,\,
\Bigg]^{2}\\
&-\left[1-\frac{3}{4\alpha r_{e}}\left(\frac{a}{b}+\frac{b}{a}-\frac{1}{\alpha r_{e}}\right)\right]\frac{l(l+1)}{(\alpha r_{e})^{2}}\Bigg\}\, .
\end{align}

By expanding  $\epsilon_{nl}$ in powers of $n+\frac{1}{2}$ and $l(l+1)$, expressions can be straightforwardly obtained for the Dunham parameters in terms of the four Pöschl-Teller parameters. 

\section{Conclusions and discussion}
The proposed approximation for the centrifugal term allows one to obtain approximately the bound states of a two-body system in which the interaction is described by a Pöschl-Teller potential. The eigenvalues are given by eq.~(\ref{epsilonnl}). The corresponding eigenfunctions are also known since the approximate radial equation for $l>0$ has the form of the one for the $s$-wave but with shifted values of the parameters $\mu$ and $\nu$.  

In contrast with the schemes based on the Greene-Aldrich approximation, the  proposed technique can be used with the full-fledged (four-parametric) Pöschl-Teller potential. 

By its nature, any Pekeris-type approximation is exact at some finite point (which is usually taken to be the equilibrium bond length) while Greene-Aldrich-type approximation schemes  are good in the limit $r\to 0$ (and, some of them, also in the limit $r\to \infty$).  When bound states are studied, a better result is to be expected from a Pekeris-type  scheme. The problem needs further investigation and it is possible that the answer will depend on the particular system considered, i.e., on the values of the Pöschl-Teller parameters.   

The second Pöschl-Teller potential has been used in the example, studied in the present letter. The same technique can be used with  the first Pöschl-Teller potential or with other exactly solvable potentials given by linear combinations of two (or more) linearly independent terms.  

\section{Acknowledgment}
The author thanks Flávio Jamil Souza Ferreira for useful discussions. 

\end{document}